\documentclass[11pt]{article}

\usepackage{epsfig}
\usepackage{amsfonts,amssymb}

\newcommand{\sect}[1]{\setcounter{equation}{0}\section{#1}}

\newcommand{\al}{\ensuremath{\alpha}}

\newcommand{\Ga}{\ensuremath{\Gamma}}

\newcommand{\ep}{\ensuremath{\epsilon}}
\newcommand{\vep}{\ensuremath{\varepsilon}}

\newcommand{\la}{\ensuremath{\lambda}}

\newcommand{\om}{\ensuremath{\omega}}
\newcommand{\Om}{\ensuremath{\Omega}}
\newcommand{\p}{\ensuremath{\phi}}
\newcommand{\s}{\ensuremath{\sigma}}
\renewcommand{\S}{\ensuremath{\Sigma}}
\renewcommand{\t}{\ensuremath{\tau}}

\newcommand{\F}{\ensuremath{{\cal F}}}

\renewcommand{\L}{\ensuremath{{\cal L}}}

\newcommand{\N}{\ensuremath{{\cal N}}}

\renewcommand{\d}{\ensuremath{{\rm d}}}

\newcommand{\ra}{\ensuremath{\rightarrow}}
\newcommand{\del}{\ensuremath{\partial}}

\newcommand{\td} {\ensuremath{\tilde}}

\newcommand{\half}{\ensuremath{\frac{1}{2}}}

\newcommand{\quarter}{\ensuremath{\frac{1}{4}}}

\newcommand{\lich}{\ensuremath{\nabla_{\rm L}}}

\newcommand{\be}{\begin{equation}}
\newcommand{\ee}{\end{equation}}

\newcommand{\ba}{\begin{eqnarray}}
\newcommand{\ea}{\end{eqnarray}}

\begin{document}

\bigskip
\hskip 4.8in\vbox{\baselineskip12pt
\hbox{hep-th/0210308} \hbox{SUSX-TH-02-025} \hbox{UUITP-14/02}}

\bigskip
\bigskip
\bigskip

\begin{center}
{\Large \bf String theory and the Classical Stability of Plane Waves}
\end{center}

\bigskip
\bigskip
\bigskip

\centerline{\bf D. Brecher$^\natural$, J. P. Gregory$^\flat$ and
 P. M. Saffin$^\sharp$}

\bigskip
\bigskip
\bigskip

\centerline{\it $^\natural$Department of Physics and Astronomy}
\centerline{\it University of British Columbia}
\centerline{\it Vancouver, British Columbia V6T 1Z1, Canada}
\centerline{\small \tt brecher@physics.ubc.ca}

\centerline{$\phantom{and}$}

\centerline{\it $^\flat$Department of Theoretical Physics}
\centerline{\it Uppsala University, Box 803, SE-751 08 Uppsala, Sweden}
\centerline{\small \tt James.Gregory@teorfys.uu.se}

\centerline{$\phantom{and}$}

\centerline{\it $^\sharp$Centre for Theoretical Physics}
\centerline{\it University of Sussex, Falmer, Brighton BN1 9QJ, U.K.}
\centerline{\small \tt mppk2@pact.cpes.susx.ac.uk}

\bigskip
\bigskip

\begin{abstract}
  \vskip 2pt
     The presence of fields with negative mass--squared typically
     leads to some form of instability in standard field theories. 
     The observation that, at least in the light--cone gauge, strings propagating in plane
     wave spacetimes can have worldsheet scalars with 
     such tachyon--like masses suggests that the supergravity
     background may itself be
     unstable.  To address this issue, we perform a perturbative
     analysis around the type IIB vacuum plane wave, the solution
     which most obviously generates worldsheet scalars with negative
     mass--squared.  We argue that this background is perturbatively
     stable.
\end{abstract}

\newpage

\baselineskip=18pt
\setcounter{footnote}{0}

%%%%%%%%%%%%%%%%%%%%%%%%%%%%%%%%%%%%%%%%%%%%%%%%%%%%%%%%%%%%%%%%%%%%%%%%%%%%%%%

\sect{Introduction}

It has been known for some time that plane--fronted, parallel
gravitational waves (pp--waves) are exact solutions of string
theory~\cite{amati:89,horowitz:90,horowitz:90b}.  The same is true of
pp--wave backgrounds with a non--trivial dilaton and null
Neveu--Schwarz--Neveu--Schwarz (NS--NS) 3--form field
strength~\cite{horowitz:90,horowitz:90b}.  These backgrounds solve the
$\beta$--function equations to all orders in $\al'$.  A particular
class of pp--wave is the exact plane wave and in recent months the
subject of plane waves in string theory has become an intense area of
research, the reasons for this renaissance being as follows: after the
discovery of maximally supersymmetric plane waves in
eleven--dimensional~\cite{kowglik:84,farrill:01} and ten--dimensional
type IIB~\cite{blau:01} supergravity, it was
observed~\cite{blau:0201,blau:0202} that these solutions can be
thought of as Penrose limits of the AdS$_p \times S^{D-p}$ vacua of
the respective theories.  Despite the fact that the ten--dimensional
case involves the Ramond--Ramond (R--R) 5--form, it was then
shown~\cite{metsaev:01,metsaev:02} that superstring theory on this
background can be solved in the light--cone gauge.  Finally, a sector
of $\N = 4$ super--Yang--Mills (SYM) which is dual to string theory on
this background was identified by Berenstein \emph{et
al}~\cite{berenstein:0202}, for the first time allowing for truly
stringy tests of, at least a special limit of, the AdS/CFT
correspondence~\cite{maldacena:97,gubser:98,witten:98}.

The Penrose limit~\cite{penrose:76} of any solution of the
ten--dimensional supergravity theories~\cite{guven:00} gives a plane
wave, which preserves at least one--half of the supersymmetries, but
sometimes more: for example, the Penrose limit of the AdS$_5 \times
T^{1,1}$ geometry, dual to a certain $\N = 1$ superconformal gauge
theory~\cite{klebanov:98}, gives the same maximally supersymmetric
plane wave~\cite{itzhaki:02, gomis:02, pando-zayas:02}.  Since this
latter work, Penrose limits have been taken of a whole host of
geometries which arise in string theory.  Of most relevance to us are
the plane waves one obtains by taking Penrose limits of the geometries
dual to \emph{non}--conformal gauge theories: the Pilch--Warner
geometry~\cite{freedman:99,pilch:0002,pilch:0006}, dual to an $\N = 1$
supersymmetric renormalization group (RG) flow between $\N = 4$ SYM
and an $\N = 1$ superconformal infra--red (IR) fixed
point~\cite{corrado:02,gimon:02,brecher:02}; the near--horizon
limits of the D$p$-brane geometries for $p \ne 3$~\cite{gimon:02}; and
various other relevant
backgrounds~\cite{blau:0202,pando-zayas:02,gursoy:02,hubeny:02,fuji:02,ryang:02}.

Issues concerning holography, even in the maximally supersymmetric
example, remain
mysterious~\cite{das:02,kiritsis:02,leigh:02,berenstein:0205,marolf:02}.
Although both this~\cite{berenstein:0205} and more general plane
waves~\cite{marolf:02}, have a boundary which is a null line, it is
not clear as to what the holographic theory actually is.  It is thus
even harder to address more general issues regarding RG flows in the
plane wave limit, though in this context they have been discussed
in~\cite{gursoy:02,das:02,hubeny:02,oz:02,fuji:02}.  Certain
general statements can be made, however: the Penrose limit of a
geometry dual to a non--conformal gauge
theory results in a plane wave with a time--dependent profile.  In
light--cone gauge string theory, it is precisely this profile which
gives a mass to the worldsheet scalars so, in this case, one
generically finds simple harmonic oscillator modes with
time--dependent masses~\cite{gimon:02}.

Moreover, the Penrose limit of the near--horizon D$p$-brane geometry
for $p \ne 3$ gives rise to plane waves with profiles of the ``wrong
sign''; that is, which generate worldsheet scalars with negative
(mass)$^2$~\cite{gimon:02}\footnote{Further examples of the occurence
of modes of negative (mass)$^2$ in Penrose limits of other
supergravity solutions are studied
in~\cite{Bhattacharya:0209054,Bhattacharya:0210072}.}.  One might
think that such terms signal instabilities although, since the
``mass'' of the worldsheet scalars is really only an artifact of the
light--cone gauge~\cite{gimon:02,cvetic:0209}, this na\"{\i}ve thought
may not be correct.  Either way, the issues involved warrant further
investigation.

Already in~\cite{brecher:02}, it was noted that, for a certain range
of parameters, one can generate what appear to be stringy
instabilities in a plane wave background with non--trivial 3--form and
5--form field strengths.  In this case, however, all the mass terms
have the correct sign and the zero modes of the string are unaffected;
one would thus not expect to see classical instabilities of the
background.  In fact, backgrounds with a non--trivial 3--form will
prove to be too complicated to analyse in any detail here.  Even in
the maximally supersymmetric case, however, the dimensionless string
theory Hamiltonian is~\cite{berenstein:0202,metsaev:02} \be H = \sum_n
\sqrt{n^2 + M^2} ~N_n,
\label{eqn:H}
\ee

\noindent where $M$ denotes the dimensionless mass of the worldsheet
scalars and $N_n$ is the occupation number for each mode at level $n$.
It would appear that switching the sign of $M^2$ will give rise to
instabilities not only for stringy modes with $n \ne 0$, but also for
the zero modes themselves\footnote{Of course, upon switching the sign
of $M^2$, one loses maximal supersymmetry and it is not obvious that
the Hamiltonian in this case is given by (\ref{eqn:H}) with a simple
sign difference.  This is one of the issues we will address.}.  And it
is precisely these latter potential instabilities which one might
expect to observe in a standard classical analysis.

Our purpose here is to assess the perturbative stability or otherwise
of plane waves.  That is, we will consider whether there exist
solutions of the linearized field equations which grow exponentially
in time.  As far as we are aware, this is the first study of classical
stability for non--static, though possibly stationary, spacetimes.  Although
we are only able to explicitly solve the relevant equations
for the plane waves with \emph{constant} profiles (the so--called
Cahen--Wallach spaces~\cite{cahen:70}), this class of spacetimes
certainly includes cases for which the (mass)$^2$ of the worldsheet
scalars is negative.  We will show that this class of plane waves does
not exhibit classical instabilities.  It would also seem that scalar
field theory in this background is well--defined, in the sense that it
too is stable~\cite{marolf:0210}.

In the following section, we briefly review pp-- and plane waves in an
arbitrary number of dimensions,
commenting on some general properties relating to the issue of
stability, including supersymmetry and particle production.  We study
the string theory dispersion relation in the
ten--dimensional vacuum plane wave in section 3, to
assess in what sense it shows signs of instabilities.  In section 4,
we compute the linearized vacuum field equations of type IIB
supergravity, specialising to the light--cone gauge to restrict
ourselves to purely physical degrees of freedom.  The resulting equations are
analysed in section 5.  Our analysis of the
metric perturbations is applicable to plane waves in any dimension,
and that of the dilaton and form--field perturbations generalize
fairly readily.  We conclude in section 6.

%%%%%%%%%%%%%%%%%%%%%%%%%%%%%%%%%%%%%%%%%%%%%%%%%%%%%%%%%%%%%%%%%%%

\sect{The spacetime}

Consider the vacuum pp--wave solution in an arbitrary number
of dimensions (in the following section we will specialize to waves of
type IIB supergravity).  The metric
\be
\d s^2 = 2 \d u \d v + H(u,x) \d u^2 + \d x^i \d x^i \qquad ( i =
1,\ldots,D-2 ),
\label{eqn:pp}
\ee

\noindent has curvature
\be
R_{uiuj} = -\half \del_i \del_j H, \qquad R_{uu} = -\half \del_i
\del_i H,
\label{eqn:curvature}
\ee

\noindent and scalar Laplacian
\be
\Box = 2 \del_u \del_v - H(u,x) \del_v^2 + \del_i\del_i,
\ee

\noindent where we should note that we will not distinguish between
upper and lower transverse indices.  In general, the only isometry of
this spacetime is that generated by the null Killing vector $\del/\del
v$.

Taking $\ep=0$ for null geodesics, $\ep=1$ for timelike geodesics, and
with $\la$ denoting the affine parameter, this isometry gives rise
to a conserved energy, $E$:
\be
u(\la) = E \la + u_0.
\label{eqn:u}
\ee

\noindent The constraint $\L = -\epsilon$ is used to solve
for $v$: with a dot denoting differentiation with respect to
$\la$, we have
\be
v(\la) = - \frac{1}{2E} \int \left( \epsilon + H(u,x) E^2 + \dot{x}^i \dot{x}^i
\right) \d \la + v_0,
\label{eqn:v}
\ee

\noindent where the transverse coordinates satisfy
\be
\ddot{x}^i = \frac{1}{2} \del_i H E^2.
\ee

\noindent For the plane wave, with $H(u,x) = -A_{ij}(u) x^i x^j$, one
can solve this equation for arbitrary initial conditions $x^i(0)$ and $\dot{x}^i(0)$.

The general pp--wave is singular, albeit in a special sense since all
curvature invariants vanish identically~\cite{horowitz:90,horowitz:90b}.  To see
this, one constructs an orthonormal frame which is parallelly
propagated along a time--like geodesic, with tangent vector
\be
T^a = \left( E, \dot{v}, \dot{x}^i \right).
\ee

\noindent The unit normals
\be
n^a_{(i)} = \left( 0, -\frac{\dot{x}^i}{E}, \delta^i_{(i)} \right),
\qquad m^a = \left( E, \dot{v} + \frac{1}{E} , \dot{x}^i \right),
\ee

\noindent all obey $\left( T \cdot \nabla n \right)^a = 0$, as
required for a parallelly propagated frame of reference.  The non--zero components
of the Riemann curvature with respect to this frame of reference,
\ba
R_{(T)(i)(T)(j)} &\equiv& R_{abcd} T^a n^b_{(i)} T^c n^d_{(j)} =
-\frac{E^2}{2} \del_i \del_j H, \\
R_{(m)(i)(m)(j)} &\equiv& R_{abcd} m^a n^b_{(i)} m^c n^d_{(j)} =
-\frac{E^2}{2} \del_i \del_j H,
\ea

\noindent will generically diverge.  If such a divergence occurs at
some point in the spacetime, then an observer will encounter infinite
tidal forces at that point, and the space is geodesically
\emph{in}complete\footnote{It is amusing to note that one encounters
these ``pp--curvature singularities'' in the solutions describing
pp--waves on fundamental
strings~\cite{kaloper:96,horowitz:97,myers:97,ross:97}, and also in
non--dilatonic ``brane--wave''
solutions of ten-- and eleven--dimensional
supergravity~\cite{brecher:00}.  In that case, even the exact plane
waves give rise to such curvature singularities at the would--be
horizon of the brane.  Presumably, the related solutions
considered in~\cite{bain:02} have a similar global structure.}.  
The only \emph{non--singular}
pp--wave~\cite{horowitz:90,horowitz:90b} is the plane wave, for which
\be
H(u,x) = -A_{ij}(u) x^i x^j,
\ee

\noindent where $A_{ij}(u)$ is a \emph{smooth} function of $u$.
(For constant $A_{ij}$, this is a Lorentzian symmetric space \cite{cahen:70,blau:01}.)
In this case,
\be
R_{(T)(i)(T)(j)} = R_{(m)(i)(m)(j)} = E^2 A_{ij}(u),
\ee

\noindent and the tidal force is attractive for positive eigenvalues
of $A_{ij}$ and repulsive for negative eigenvalues.  It might appear
that this notion of a singularity is observer--dependent, but this is
not the case since the tidal force, divergent or otherwise, would be
experienced by \emph{all} time--like observers.  Indeed, one can show
that the metric function $A_{ij} (u)$ encodes diffeomorphism--invariant
properties of the spacetime~\cite{marolf:0210}.

The nature of the tidal force is seen in the geodesics also.  The
vacuum field equation is simply
\be
{\rm tr} A (u) = 0,
\ee

\noindent and, for constant $A_{ij}$, we can always diagonalize,
taking $A_{ij} = \la_{(i)} \delta_{ij}$.  Then there must be at least
one negative eigenvalue, and we can write $\la_{(i)} = \pm
\mu^2_{(i)}$.  The geodesic equations for this simple case are easily
solved: $u$ and $v$ are as in (\ref{eqn:u}) and (\ref{eqn:v}), and \be
x^i = A^i \cos ( \mu_{(i)} E \la ) + B^i \sin ( \mu_{(i)} E \la )
\qquad ( \la_{(i)} = \mu^2_{(i)} ),
\label{eqn:geo1}
\ee
\be
x^i = A^i \exp ({\mu_{(i)} E \la}) + B^i \exp (
-\mu_{(i)} E \la ) \qquad ( \la_{(i)} = -\mu^2_{(i)} ).
\label{eqn:geo2}
\ee

\noindent The attractive tidal force in the first case gives
oscillatory behaviour around the origin, and the repulsive tidal force
in the latter case gives geodesics which are pushed off to infinity.
In this sense, the vacuum spacetime exhibits an instability (and we
will see the same behaviour in the zero modes of the string in this
background below).  This behaviour is reminiscent of de Sitter space,
in which particles also accelerate forever and, just as in that case,
the repulsive tidal force in the directions associated with negative
eigenvalues does not necessarily give rise to instabilities in the
usual sense\footnote{The Penrose limit of pure de Sitter space is just
flat space, as for pure anti--de Sitter space~\cite{blau:0202}.
Although such a spacetime is not a solution of any known supergravity
theory, if we consider the Penrose limit of de Sitter space times a
sphere, along a geodesic which has angular momentum on the sphere,
then we do find negative eigenvalues coming from the de Sitter
directions.}.  In the case of de Sitter space, stability is guaranteed
by the ``cosmic no hair''
conjecture~\cite{gibbons:77,hawking:82,wald:82}, and one might wonder
if there is a similar argument for the case of the plane wave.

Since all such plane waves are at least one--half supersymmetric, one
might be tempted to argue that they \emph{must} be stable.  However,
the extent to which the standard
arguments~\cite{witten:81,gibbons:82,gibbons:83,gibbons:93} concerning
stability and supersymmetry apply here is unclear.  After all, the
spacetime is not asymptotically flat.  Moreover, the supersymmetries
common to all such plane waves do not commute with the Hamiltonian.
One can thus write down one--half supersymmetric configurations which
nevertheless do \emph{not} solve the relevant field
equations~\cite{farrill:99}.  However, demanding extra or, in the
language of~\cite{cvetic:0203082,cvetic:0203229}, supernumerary,
supersymmetries does restrict one to solutions of the field equations;
and indeed it would seem that any such solution which preserves more
than one--half of the supersymmetries does not give rise to the
potential problems alluded to above.  For example, it is fairly easy
to see that waves with time--dependent profiles never have such
supernumerary supersymmetries.

More explicitly, the integrability condition for the
existence of Killing spinors is
\be
[D_u,D_i] \ep = \quarter \del_i \del_j H ~\Ga_j \Ga_+ \ep = 0,
\ee

\noindent and multiplying this from the left with $\Ga_i$ gives
\be
\del_i \del_i H ~\Ga_+ \ep = 0,
\ee

\noindent where $\del_i \del_i H = 0$ is the field equation.  One can
thus have the standard supersymmetries, satisfying $\Ga_+ \ep = 0$,
without solving the field equation.  Supernumerary supersymmetries,
however, have $\Ga_+ \ep \ne 0$ so, if any of these are to exist, the
field equation must be satisfied.  Of course, in this vacuum example,
there are no supernumerary supersymmetries anyway.  In the more
general case with background fluxes, it is true that the existence of
supernumerary supersymmetries implies the field equations, but this is
harder to see.  One has to consider the Killing spinor equation
directly, in addition to the variation of any other fermionic fields,
for example the dilatino in type IIB supergravity.  An analogous study
of the integrability condition for the existence of Killing spinors in
this more general case is not enough\footnote{We thank the anonymous
referee for pointing out the problems with our original argument in
this case.}.

Further general comments are in order with respect to particle production
in these backgrounds.  It is easy to see that Gibbons' original
argument~\cite{gibbons:75} --- that there is no production of massless
scalars in the sandwich wave --- can be generalized to \emph{any}
pp--wave (see, \emph{e.g.},~\cite{horowitz:90, horowitz:90b}).  Given that
$\del/\del v$ generates an isometry, we can always decompose the modes
of a scalar field in the pp--wave background with respect to $v$, as
\be \p (u,v,x) = e^{ip_-v} \td{\p} (u,x).  \ee

\noindent To determine whether there is particle production, we need
to consider whether the ``in'' modes in the asymptotic past are
orthogonal to the ``out'' modes in the asymptotic future\footnote{In
the case of the sandwich wave, in the asymptotic past and future, the
spacetime is flat and the obvious ``in'' and ``out'' states can be
constructed~\cite{gibbons:75}.  More generally, one can
imagine~\cite{gimon:02} an RG flow from some constant $A_{ij}$ to some
other constant $A_{ij}$ in which case one can define ``in'' and
``out'' states in these stationary regions of spacetime.}.  The
Klein--Gordon scalar product is \be (\p,\psi) = -i \int_{\S} \d \S^a
\sqrt{g_{\S}} ~\p \stackrel{\leftrightarrow}{\del_a} \psi^*,
\label{eqn:kg}
\ee

\noindent where $\S$ is the null surface $u=0$, with unit normal
$\del/\del v$.  Despite the fact that these spacetimes are not
globally hyperbolic~\cite{penrose:65} (see also~\cite{marolf:02}),
this is a good substitute for a Cauchy
surface~\cite{gibbons:75}: it fails to capture only those geodesics moving
parallel to the wave, with $u=\mbox{constant}$, and these will never
interact with the wave itself.  Explicitly, we have~\cite{gibbons:75}
\be
(\p,\psi) = -i \int_{\S} \d v \d^{D-2} x ~\p
\stackrel{\leftrightarrow}{\del_v} \psi^*,
\ee

\noindent so the mode functions are always orthogonal,
since they are simple exponentials in $v$.  For any pp--wave, the null
Killing vector $\del/\del v$ allows for a preferred notion of
frequency, so there can be no particle production and the resulting Bogoliubov
coefficients must vanish.

One might then argue that scalar fields in the pp--wave background are always
stable, since presumably an instability in such a field will cause
runaway particle production.  What is less clear is whether this
implies stability of the background itself (some comments
in~\cite{gibbons:86} seem to be make this connection).  One should
also consider what effects interactions have~\cite{marolf:0210}.

%%%%%%%%%%%%%%%%%%%%%%%%%%%%%%%%%%%%%%%%%%%%%%%%%%%%%%%%%%%%%%%%%%%

\sect{String theory on the plane wave}

Of course, the zero modes of a string in the plane wave background
will follow the same geodesics as above but we should also consider
the non--zero modes to see what effect, if any, they have.  Moreover,
it would appear from (\ref{eqn:H}) that the zero modes themselves will
give rise to instabilities for negative eigenvalues of the metric
function $A_{ij}$, and it is this issue that we want to address here.
Ultimately, we are interested in whether or
not there are tachyonic modes of the string (or particle) in the plane
wave background.  Since the notion of ``mass'' in this case is not the
same as in flat space, we will ask the question as to whether there is
faster--than--light propagation of any modes.  It would appear that there
\emph{is} though, again, in a similar way as for de Sitter space.  

Since we are still working with the vacuum solution, the worldsheet
fermions are unchanged from the usual flat space results.  In the
light--cone gauge, they are
affected only by background fluxes~\cite{metsaev:02}.  Indeed, since
the vacuum backgrounds are necessarily only one--half supersymmetric,
there are no linearly realized worldsheet supersymmetries at all ---
after fixing the kappa--symmetry, only supernumerary supersymmetries
are linearly realized~\cite{cvetic:0209}.  Then it should not be
surprising that, whereas the worldsheet scalars are massive, the
fermions will still be massless.  We will thus consider only bosonic
string theory here.

The action
\be
S = -\frac{1}{4\pi\al'} \int \d \s \d \t \sqrt{-h} h^{\al\beta}
\del_{\al} X^a \del_{\beta} X^b g_{ab},
\ee

\noindent with background metric
\be
\d s^2 = 2 \d u \d v - A_{ij}(u) x^i x^j \d u^2 + \d x^i \d x^i \qquad
(i = 1,\ldots,8),
\ee

\noindent has the energy--momentum tensor
\be
T_{\al\beta} = -\frac{1}{\al'} \left( \del_{\al} X^a \del_{\beta}
X^b g_{ab} - \frac{1}{2} h_{\al\beta} \del X^a \cdot \del X^b g_{ab}
\right),
\ee

\noindent and canonical momenta
\ba
P_- &=& \frac{\del \L}{\del \dot{V}} = \frac{1}{l_s^2} \dot{U}, \\
P_+ &=& \frac{\del \L}{\del \dot{U}} = \frac{1}{l_s^2} \left(
\dot{V} - A_{ij} X^i X^j \dot{U} \right), \\
P_i &=& \frac{\del \L}{\del \dot{X}^i} = \frac{1}{l_s^2} \dot{X}^i.
\ea

\noindent where $l_s^2 = 2 \pi\al'$.  The $V$ equation implies that
$\Box U = 0$, so taking $U=U(\t)$, we have \be U = l_s^2 P_- \t + U_0,
\ee

\noindent which is the light--cone gauge.  In that case, the
constraint $T_{00} = 0$ gives, upon substitution of the momenta,
\be
-2 P_+ P_- = P_-^2 A_{ij} X^i X^j + P_i P_i + \frac{\del_\s X^i \del_\s
X^i}{l_s^4}.
\label{eqn:T00}
\ee

$V$ is given implicitly by the constraint,
leaving the transverse coordinates as the physical degrees of
freedom.  Taking the metric function $A_{ij} = \la_{(i)}
\delta_{ij}$, and defining the dimensionless mass $M_{(i)} =
\mu_{(i)} l_s^2 P_-$, we have
\be
\Box X^i \mp M^2_{(i)} X^i = 0 \qquad (\la_{(i)} = \pm \mu_{(i)}^2).
\ee

\noindent Consider, first, the zero modes:
\ba
&& X_0^i (\t) = x^i \cos \left( M_{(i)} \t \right) + \frac{\al'}{M_{(i)}}
\sin \left( M_{(i)} \t \right) p^i \qquad (\la_{(i)} = \mu_{(i)}^2), \\
&& X_0^i (\t) = x^i \cosh \left( M_{(i)} \t \right) + \frac{\al'}{M_{(i)}}
\sinh \left( M_{(i)} \t \right) p^i \qquad (\la_{(i)} = -\mu_{(i)}^2),
\ea

\noindent which of course match the geodesics (\ref{eqn:geo1}) and
(\ref{eqn:geo2}) above.  The centre--of--mass of the string
will thus oscillate about the origin of those directions associated
with positive eigenvalues, and will accelerate forever in the
directions associated with negative eigenvalues.

The non--zero modes are solved for by making the Ansatz
\be
X^i (\t,\s) = \sum_{n\ne 0} C^i_n e^{i(\om_n \t + n \s)},
\ee

\noindent giving
\be
\om_n^2 = n^2 \pm M_{(i)}^2 \qquad (\la_{(i)} = \pm \mu_{(i)}^2),
\label{eqn:mode}
\ee

\noindent For negative eigenvalues, the set of modes with $|n| <
M_{(i)}$ have $\om_n^2 < 0$.  As in the more general case considered
in~\cite{brecher:02}, some of the $\om_n$ become imaginary, and would
seem to correspond to unstable modes.  However, this is an instability
with respect to mode creation on the string worldsheet, in precisely
the sense of~\cite{horowitz:90,horowitz:90b}, and not in the sense
that we are concerned with here.  We do not expect to see truly
stringy instabilities in our perturbative analysis.  Again as
in~\cite{brecher:02}, these stringy instabilities exist only for some
range of $n$.  In particular, we have unstable modes only if \be
\mu_{(i)} P_- > T, \ee

\noindent $T = 1/l_s^2$ being the string tension.  Unstable modes only
appear if either the light--cone momentum of the string, or the
curvature of the background, is large with respect to the string
scale (the same is true in the more
general case considered in~\cite{brecher:02}).

In the directions associated with positive eigenvalues, the oscillator
modes are given by~\cite{metsaev:02}
\be
X^i = i \sqrt{\frac{\al'}{2}} \sum_{n\ne 0} \frac{1}{\om_n} \left( \al^i_n
e^{in\s} + \td{\al}^i_n e^{-in\s} \right) e^{-i\om_n \t},
\ee

\noindent where, for $n>0$, $\om_n = \sqrt{n^2 + M_{(i)}^2}$ and, to
ensure reality, \be \om_{-n} = - \om_n, \qquad
\left(\al^i_n\right)^\dagger = \al^i_{-n}, \qquad
\left(\td{\al}^i_n\right)^\dagger = \td{\al}^i_{-n}.
\label{eqn:reality}
\ee

\noindent In the remaining directions, associated with negative
eigenvalues, we have \be X^i = i \sqrt{\frac{\al'}{2}} \sum_{|n| >
M_{(i)}} \frac{1}{\om_n} \left( \al^i_n e^{in\s} + \td{\al}^i_n
e^{-in\s} \right) e^{-i\om_n \t} + \sqrt{\frac{\al'}{2}} \sum_{|n| <
M_{(i)}} \frac{1}{\Om_n} \left( \beta^i_n e^{in\s} +
\left(\beta^i_n\right)^\dagger e^{-in\s} \right) e^{\Om_n \t}, \ee

\noindent where, for $n>0$, $\om_n = \sqrt{n^2 - M_{(i)}^2}$, and the
same conditions as in (\ref{eqn:reality}) apply to ensure reality.
The $\beta^i_n$ operators are associated with the exponentially
growing/decaying modes and we have set $\om_n = i \Omega_n$ for $|n| <
M_{(i)}$ where, for $n>0$, $\Omega_n = \sqrt{M_{(i)}^2 - n^2}$.  Then
the expansion for the unstable modes is manifestly real.  To include
all the relevant modes in this case, we also set $\Omega_{-n} =
-\Omega_n$.

The non--vanishing Poisson brackets are then
\[
\left[ x^i, p^j \right]_{PB} = \delta^{ij},
\]
\[
[\al^i_m,\al^j_n]_{PB} = [\td{\al}^i_m,\td{\al}^j_n]_{PB} = -i \om_m
\delta_{m+n,0} \delta^{ij},
\]
\be
[\beta^i_m,\beta^j_n]_{PB} = [\left(\beta^i_m\right)^\dagger,\left(\beta^j_n\right)^\dagger]_{PB} =  \Om_m
\delta_{m+n,0} \delta^{ij}.
\ee

\noindent Upon quantisation, the modes with real $\om_n$ will give
rise to the usual harmonic oscillators, but those with imaginary
$\om_n$ have an extra factor of $i$.  In other words, the $\beta^i_n$ do not
correspond to harmonic oscillators, rather they are associated with
the unstable worldsheet modes.

Substituting for the mode expansions of the worldsheet scalars and
the respective momenta into (\ref{eqn:T00}) gives
\[
-2(2\pi)^2 P_+ P_- = (2\pi)^2 P_-^2 \la_{(i)} x^i x^i + p^i p^i +
\frac{1}{\al'} \sum_{n\ne 0} \left( \al^i_n \al^i_{-n} +
\td{\al}^i_n \td{\al}^i_{-n} \right)
\]
\be
+ \frac{1}{\al'} \sum_{|n| < M_{(i)}} \left( \beta^i_n \beta^i_{-n} +
\left(\beta^i_n\right)^\dagger \left(\beta^i_{-n}\right)^\dagger \right),
\ee

\noindent where the first oscillator sum runs over all directions
associated with positive eigenvalues, and over those modes for which
$|n|>M_{(i)}$ in the directions associated with negative eigenvalues.
The second sum applies to the latter directions only.  The $P_-^2$
contribution comes from the zero modes, so would also be present in
the case of a particle.  In either case, it is not at all obvious as
to how this dispersion relation gives us information on what our
flat--space intuition would call a ``mass''.  In flat space, we have
\be
-2 P_+ = \frac{1}{P_-} \left( p^i p^i + \mbox{oscillators}
\right),
\ee

\noindent which gives well--defined behaviour; in particular, $P_+ \ra
0$ as $P_- \ra \infty$, so that we never have faster--than--light
propagation.  In the case at hand, however, we have
\be
-2 P_+ = P_-
\la_{(i)} x^i x^i + \frac{1}{P_-} \left( p^i p^i + \mbox{oscillators}
\right),
\ee

\noindent the non--standard dependence on $P_-$ giving rise to
``unstable'' behaviour: depending on the sign of $\la_{(i)}$, we have $P_+
\ra \pm \infty$ as either $P_- \ra \infty$ or $\mu_{(i)} \ra \infty$.
In either case, we know that in this limit, only the zero modes
survive, since all the stringy modes with $n>0$ are unstable; the
string is literally ripped apart.  Note that modes with small $P_-$
behave as if they have a positive mass, so in this infra--red regime,
there are no tachyonic excitations.  

At any rate, based on flat--space intuition it would appear that we
either have slower--than--light or faster--than--light propagation
depending on this sign.  Negative eigenvalues seem to give tachyonic
behaviour, but this is presumably to be expected since particles (and
strings) accelerate forever.  Even so, we will not find any unstable
modes in the analysis below.  As we will see, the problem in this
simple case reduces to the quantum mechanics of the inverted
oscillator in which similar behaviour is observed: quantum particles
accelerate forever, but the system is perturbatively stable.  Various
subtleties concerning this dispersion relation are addressed
in~\cite{marolf:0210}, where it is argued that imposing boundary
conditions at some finite $x^i x^i = L^2$ effectively removes the
problems associated with large $P_-$.  

To make contact with (\ref{eqn:H}), we rescale
\be
\al^i_n = \sqrt{\om_n} a^i_n, \qquad \al^i_{-n} = \sqrt{\om_n}
\bar{a}^i_n, \qquad \beta^i_n = \sqrt{\Om_n}, \qquad \beta^i_{-n} =
\sqrt{\Om_n}\bar{b}^i_n,
\ee

\noindent and likewise for the $\td{\al}$ and $\beta^\dagger$, and
combine the zero modes as
\be
a^i_0 = \frac{1}{\sqrt{M_{(i)} \al'}} \left( \al' p^i - i M_{(i)} x^i \right),
\qquad \bar{a}^i_0 = \frac{1}{\sqrt{M_{(i)}\al'}} \left( \al' p^i + i
M_{(i)} x^i \right).
\ee

\noindent Upon quantisation, the new operators obey
\be
[a^i_0,\bar{a}^j_0]=\delta^{ij}, \qquad
[a^i_m,\bar{a}^j_n]=\delta_{mn} \delta^{ij}, \qquad
[b^i_m,\bar{b}^j_n]=i\delta_{mn}\delta^{ij},
\ee

\noindent and likewise for the $\td{\al}$ and $\beta^\dagger$, and we have
\be
-2P_+P_- = \frac{1}{\al'} \sum_i \left( M_{(i)} N_0^{(i)} + \sum_{n>0}
\om_n^{(i)} N_n^{(i)} + \sum_{|n| < M_{(i)}} \Om_n^{(i)}
\hat{N}_n^{(i)} \right), 
\ee

\noindent where $M_{(i)} = \pm \mu_{(i)} l_s^2 P_-$, $\om_n^i =
\sqrt{n^2 + M_{(i)}^2}$, $N_n^{(i)} = a^i_n \bar{a}_n^i + \td{a}^i_n
\td{\bar{a}}^i_n$ and $\hat{N}_n^{(i)} = b^i_n \bar{b}^i_n +
(b^i_n)^\dagger (\bar{b}^i_n)^\dagger$.

%%%%%%%%%%%%%%%%%%%%%%%%%%%%%%%%%%%%%%%%%%%%%%%%%%%%%%%%%%%%

\sect{Linearized field equations}

We will consider perturbations of the type IIB version of the vacuum plane
wave discussed in section 2.  We have some results regarding the
non--vacuum solution, but will not discuss them here since this case
is much harder to analyse in any detail.
Moreover, the vacuum solution is the cleanest in which possible
negative (mass)$^2$ terms are generated on the string worldsheet.

The field content of the bosonic sector of type IIB supergravity
is: a complex scalar, $\Phi = \chi + i e^{-\p}$, combining
the axion and dilaton, $\chi$ and $\p$; a complex 2--form potential
with field strength
\be
G_3 = \d A_2 = H_3 + i F_3,
\ee

\noindent combining the NS--NS and R--R 3--form field strengths, $H_3$ and
$F_3$; and an R--R 4-form potential with self--dual field strength
\be
F_5 = \star F_5 = \d C_4 - \frac{1}{8} \Im \left( A_2 \wedge G_3^*
\right),
\ee

\noindent where $*$ denotes complex conjugation and $\star$ denotes
ten--dimensional Hodge duality.  We take the field equations
from~\cite{schwarz:83}, perturbations of which, in a vacuum
background, simplify dramatically.

\subsection{Scalar perturbations}

Throughout the analysis we will find ourselves solving the field
equations similar to those for a massless scalar, so consider
the complex scalar, $\Phi$, which satisfies
\be
\Box \Phi = 0.
\ee

\noindent We want to look for instabilities, \emph{i.e.} modes which grow
exponentially in time, where ``time'' is $u$.  We will also demand
that the modes are normalizable --- either $L^2$ normalizable or
delta--function normalizable --- in $v$ and all transverse directions.  We have
already discussed above that we can always decompose the modes with
respect to the Killing vector $\del/\del v$, so we make the Ansatz
\be
\Phi = e^{i(p_- v - p_+ u)} \F(x),
\ee

\noindent where $\F(x) = \p_1(x) + i \p_2(x)$ depends on all
transverse coordinates.  Then
\be
\Box \Phi = e^{i(p_- v - p_+ u)} \left[ (2p_-p_+ + H(u,x) p_-^2) \F(x)
+ \del_k \del_k \F(x) \right].
\ee

\noindent We assume that $p_-$ is real, else the modes will not be
normalizable in the $v$ direction, and unstable modes will have imaginary
$p_+$.   Our approach will be to consider those modes with real $p_+$
and show that in fact this includes all modes with real momentum
$p_-$, so \be [2p_-p_+ + H(u,x) p_-^2] \p(x) + \del_k \del_k \p(x) = 0,
\label{eqn:equation}
\ee

\noindent for both $\p=\p_1$ and $\p=\p_2$.  We will see that precisely the
same equation controls the perturbations of all supergravity fields, and will
analyse the solutions in the following section.

\subsection{Metric perturbations}

As usual in perturbation theory, we need to choose a gauge to restrict
ourselves to physical degrees of freedom.  Although the standard
transverse, trace--free gauge considerably simplifies the linearized
field equations, this choice does not entirely fix the gauge.  We will
rather make use of the light--cone gauge, which is well--suited to our
background spacetime; but also has the advantage of fixing the gauge
freedom entirely.  The field equation\footnote{One often thinks of the
linearized vacuum field equation to be the vanishing of the
Lichnerowicz operator~\cite{lich:61} acting on the metric
perturbations.  This is true only in the transverse, trace--free gauge
however.  That is $\lich h_{ab} = -2\delta R_{ab}$ only if $h =
g^{ab}h_{ab} = \nabla_b h^b_a = 0$.  Since we will not be working in
this gauge, our vacuum field equation is $\delta R_{ab} = 0$, and not
$\lich h_{ab} = 0$.} is $\delta R_{ab} = 0$.

The light--cone gauge is defined by
\be
h_{va} = 0,\qquad\forall\;\;a,
\label{eqn:lcg}
\ee

\noindent and substituting for this gives
\ba
-2 \delta R_{uu} &=& \Box h_{uu} - 2\del_u ( \del_v h_{uu} + \del_i h_{iu}
) - \del_i \del_j H h_{ij} - 2\del_i H \del_j h_{ij}, \\
-2 \delta R_{uv} &=& -\del_v ( \del_v h_{uu} + \del_i h_{iu} ), \\
-2 \delta R_{ui} &=& \Box h_{ui} + \del_j H \del_v h_{ij} - \del_i (
\del_v h_{uu} + \del_j h_{ju} ), \\
-2 \delta R_{vv} &=& \del_v^2 h_{ii}, \\
-2 \delta R_{vi} &=& -\del_v ( \del_v h_{iu} + \del_j h_{ij} ), \\
-2 \delta R_{ij} &=& \Box h_{ij}.
\ea

\noindent For the maximally supersymmetric plane wave, these results
were effectively derived in~\cite{metsaev:02,das:02}.  As in that
case, the equations $\delta R_{va} = 0$ appear as constraints, which
we solve by taking \be h_{ii} = 0, \qquad \del_v h_{iu} + \del_j
h_{ij} = 0, \qquad \del_v h_{uu} + \del_i h_{iu} = 0,
\label{eqn:integrate}
\ee

\noindent which leaves the equations
\be
\Box h_{ij} = 0, \label{eqn:h0}
\ee
\be
\Box h_{iu} + \del_j H \del_v h_{ij} = 0, \label{eqn:h1}
\ee
\be
\Box h_{uu} - \del_i \del_j H h_{ij} - 2 \del_i H \del_j h_{ij} =
0. \label{eqn:h2}
\ee

\noindent We thus need only solve
\be
\Box h_{ij} = 0,
\ee

\noindent all other components of $h_{ab}$ following from $h_{ij}$
\emph{via} the constraints (\ref{eqn:integrate}).  Working in the light--cone
gauge reduces the degrees of freedom to the physical ones only, the
transverse components $h_{ij}$, minus the trace.

We make the Ansatz
\be
h_{ij} = \Re \left\{ \xi_{ij} ~e^{i(p_- v - p_+ u)} \F(x) \right\},
\label{eqn:metric_ansatz}
\ee

\noindent where, as in the previous subsection, $\F(x) = \p_1(x) + i
\p_2(x)$.  $\xi_{ij} = \xi_{(ij)}$ denotes a constant polarisation
tensor which satisfies $\xi_{ii} = 0$. Again we consider modes
normalizable in $v$ (real $p_-$) with real $p_+$.  Then \be \Box
h_{ij} = \Re \left\{ \xi_{ij} e^{i(p_- v - p_+ u)} \left[ (2p_-p_+ +
H(u,x) p_-^2) \F(x) + \del_k \del_k \F(x) \right] \right\} = 0.  \ee

\noindent Writing this in terms of real and imaginary parts, we have
\be
[2p_-p_+ + H(u,x) p_-^2] \p(x) + \del_k \del_k \p(x) = 0,
\ee

\noindent for both $\p=\p_1$ and $\p=\p_2$, just as in
(\ref{eqn:equation}) above.\

The constraints (\ref{eqn:integrate}) give \be h_{iu} = \Re \left\{
\frac{i}{p_-} \xi_{ij} ~e^{i(p_- v - p_+ u)} \del_j \F(x) \right\},
\qquad h_{uu} = - \Re \left\{ \frac{1}{p_-^2} \xi_{ij} ~e^{i(p_- v -
p_+ u)} \del_i \del_j \F(x) \right\}, \ee

\noindent though we still need to check for consistency with
the field equations (\ref{eqn:h1}) and (\ref{eqn:h2}).  Substituting
for the above results, we find
\ba
(4.16) &\Rightarrow& \Re \left\{ \frac{i}{p_-}
\xi_{ij} ~e^{i(p_- v - p_+ u)} \del_j \left[(2p_-p_+ + H(u,x) p_-^2)
\F(x) + \del_k \del_k \F(x) \right] \right\} = 0,\\
(4.17) &\Rightarrow&
\Re \left\{ \frac{1}{p_-^2}
\xi_{ij} ~e^{i(p_- v - p_+ u)} \del_i \del_j \left[(2p_-p_+ + H(u,x) p_-^2)
\F(x) + \del_k \del_k \F(x) \right] \right\} = 0,
\ea

\noindent so consistency is guaranteed and we need only take $\p_1$
and $\p_2$ to satisfy (\ref{eqn:equation}).  Solutions of this
equation thus fix \emph{all} metric perturbations.

\subsection{4--form perturbations}

In addition to (\ref{eqn:lcg}), we now also impose
\be
\delta C_{vabc}=0,\qquad\forall\;\;a,b,c,
\ee

\noindent so that
\be
\delta F_{v a_1 \ldots a_4} = \del_v \delta C_{a_1 \ldots a_4}.
\ee

\noindent Perturbing the self--duality condition $F_5 = \star
F_5$ gives
\ba
&& \delta F_{v i_1 \ldots i_4} = -\frac{1}{4!} \vep_{i_1 \ldots i_4 j_1
\ldots j_4} \delta F_{v j_1 \ldots j_4}, \label{eqn:f1} \\
&& \delta F_{uv ijk} = \frac{1}{5!} \vep_{ijk i_1 \ldots i_5} \delta
F_{i_1 \ldots i_5}, \label{eqn:f2} \\
&& \delta F_{u i_1 \ldots i_4} = \frac{1}{4!} \vep_{i_1 \ldots i_4 j_1
\ldots j_4} \delta F_{u j_1 \ldots j_4},
\ea

\noindent where $\vep_{i_1 \ldots i_8} = \vep_{uv i_1 \ldots i_8}$.
The 4--form potential is anti--self--dual in eight dimensions, since
(\ref{eqn:f1}) gives
\be
\delta C_{i_1 \ldots i_4} = -\frac{1}{4!} \vep_{i_1 \ldots i_4 j_1 \ldots
j_4} \delta C_{j_1 \ldots j_4}.
\label{eqn:asd}
\ee

\noindent Furthermore, (\ref{eqn:f2}) is just the constraint
\be
\del_v \delta C_{uijk} + \del_l \delta C_{lijk} = 0,
\label{eqn:self-dual}
\ee

\noindent which can be solved for $\delta C_{uijk}$.

The 4--form equation of motion is
\be
\nabla^e \delta F_{eabcd} = 0,
\ee

\noindent the $\{u,v,i,j\}$ component of which can be solved by taking
\be
\del_k \delta C_{kuij} = 0,
\label{eqn:4--form1}
\ee

\noindent and the $\{v,i,j,k\}$ component gives the constraint
(\ref{eqn:self-dual}).  The remaining components are
\be
\Box \delta C_{ijkl} = 0,
\label{eqn:4--form2}
\ee
\be
\Box \delta C_{uijk} + \del_l H \del_v \delta C_{lijk} = 0,
\label{eqn:4--form3}
\ee

\noindent and, again, we will see below that the solution for $\delta
C_{uijk}$ found from the constraint (\ref{eqn:self-dual}) is
consistent with this field equation.

Since the 4--form is real, we make a similar Ansatz to
(\ref{eqn:metric_ansatz}) above:
\be
\delta C_{ijkl} = \Re \left\{ \ep_{ijkl} ~e^{i(p_- v - p_+ u)} \F(x)
\right\},
\ee

\noindent where
\be
\ep_{i_1 \ldots i_4} = -\frac{1}{4!} \vep_{i_1 \ldots i_4 j_1 \ldots j_4}
\ep_{j_1\ldots j_4},
\ee

\noindent is a constant polarisation tensor, anti--self--dual in eight
dimensions, and $\F(x) = \p_1 (x) + i \p_2(x)$.  Then
\be
\Box \delta
C_{ijkl} = \Re \left\{ \ep_{ijkl} ~e^{i(p_- v - p_+ u)} \left[
(2p_-p_+ + H(u,x) p_-^2) \F(x) + \del_m \del_m \F(x) \right] \right\}
= 0.
\ee

\noindent The constraint (\ref{eqn:self-dual}) gives
\be
\delta C_{uijkl} = \Re \left\{ \frac{i}{p_-} \ep_{lijk} ~e^{i(p_- v -
p_+ u)} \del_l F(x) \right\},
\ee

\noindent so that (\ref{eqn:4--form1}) is an identity, and
substituting this into the field equation (\ref{eqn:4--form3}) gives
\be
\Re \left\{
\frac{i}{p_-} \ep_{lijk} ~e^{i(p_- v - p_+ u)} \del_l \left[
(2p_-p_+ + H(u,x) p_-^2) \F(x) + \del_m \del_m \F(x) \right] \right\} =
0.
\ee

\noindent Again, both $\p=\p_1$ and $\p=\p_2$ must satisfy
(\ref{eqn:equation}).

\subsection{2--form perturbations}

The 2--form perturbations go through in the same way as the for the
4--form above, the only difference being that they are complex.  Taking
\be
\delta A_{va} = 0,\qquad\forall\;\;a
\ee

\noindent gives
\be
\delta G_{vab} = \del_v \delta A_{ab}.
\ee

\noindent The equation of motion is
\be
\nabla^c \delta G_{cab} = 0,
\ee

\noindent the $\{u,v\}$ and $\{v,i\}$ components of which are
constraints which can be solved by taking
\be
\del_i \delta A_{iu} = 0, \qquad \del_v \delta A_{ui} + \del_j \delta
A_{ji} = 0.
\label{eqn:3--form_constr}
\ee

\noindent This leaves
\be
\Box \delta A_{ij} = 0,
\ee
\be
\Box \delta A_{ui} + \del_j H \del_v \delta A_{ji} = 0,
\label{eqn:3--form_eqn}
\ee

\noindent the structure of which is the same as for the 4--form
perturbations.  Writing
\be
\delta A_{ij} = \zeta_{ij} ~e^{i(p_- v - p_+ u)} \F(x),
\ee

\noindent for some constant complex polarisation tensor $\zeta_{ij} =
\zeta_{[ij]}$, gives
\be
\Box \delta A_{ij} = \zeta_{ij} ~e^{i(p_- v - p_+ u)} \left[
(2p_-p_+ + H(u,x) p_-^2) \F(x) + \del_k \del_k \F(x) \right] = 0.
\ee

\noindent The constraint (\ref{eqn:3--form_constr}) gives
\be
\delta A_{ui} = \frac{i}{p_-} \zeta_{ji} ~e^{i(p_- v - p_+ u)} \del_j
\F(x),
\ee

\noindent and the field equation (\ref{eqn:3--form_eqn}) is
\be
\frac{i}{p_-}
\zeta_{ji} ~e^{i(p_- v - p_+ u)} \del_j \left[
(2p_-p_+ + H(u,x) p_-^2) \F(x) + \del_k \del_k \F(x) \right] = 0,
\ee

\noindent so we see precisely the same structure as above.

%%%%%%%%%%%%%%%%%%%%%%%%%%%%%%%%%%%%%%%%%%%%%%%%%%%%%%%%%%%%%%%%%%%%%

\sect{Classical stability}

In the above section, we showed that the linearized field equations
effectively reduce to the problem of solving (\ref{eqn:equation}):
\be
[2p_-p_+ + H(u,x) p_-^2] \p(x) + \del_k \del_k \p(x) = 0.
\ee

\noindent Solutions of this equation control the
perturbations of all supergravity fields.  So far we have considered those modes with
real momentum $p_-$ that
also have real $p_+$. Here we will see that in fact \emph{all} modes
of real $p_-$ have real $p_+$. This is done by finding the general
solution to (\ref{eqn:equation}) which, using a separation of variables,
comes down to a second order o.d.e. with two known, linearly independent
solutions.  Note that we have ignored an arbitrary function of $u$
which could be added to the Ans\"{a}tze above.  That is, we could have
taken
\be
\Phi = e^{i(p_- v - p_+ u)} \F(x) + f(u),
\ee
\noindent and likewise for the other supergravity fields.
This is a manifestation of the fact that \mbox{$u={\rm
constant}$} is not a Cauchy surface, but a null surface.
However, treating it as a substitute Cauchy surface only
misses those modes which travel along constant $u$
\cite{gibbons:75}, parallel to the wave front.
The presence of such a function $f(u)$
is familiar in light-front studies \cite{heinzl:98} and
is determined by the boundary condition that fields should
vanish asymptotically, requiring $f(u)=0$.

Let us consider the simplest case, taking $A_{ij} = \la_{(i)}
\delta_{ij}$ with the eigenvalues $\la_{(i)} =
\mu^2_{(i)}, i=1, \ldots ,n$ and $\la_{(i)} = - \mu_{(i)}^2, i=n+1,
\ldots ,D-2$.   Separating variables as
\be
\F(x) = F_1(x^1) \ldots F_n(x^n)G_1(y^1)\ldots G_{D-2-n}(y^{D-2-n}),
\label{eqn:ansatz}
\ee

\noindent where we denote the directions associated with negative
eigenvalues by $\{y^1, \ldots, y^{D-2-n} \}$, gives
\be
F_{(i)}'' + \left( E_{(i)} - \mu^2_{(i)} p_-^2 (x^i)^2 \right) F_{(i)}
= 0, \qquad G_{(i)}'' + \left( E_{(i)} + \mu_{(i)}^2 p_-^2 (y^i)^2
\right) G_{(i)} = 0,
\label{eqn:sep}
\ee

\noindent where
\be
\sum_{i=1}^{D-2} E_{(i)} = p_+ p_-.
\label{eqn:Esum}
\ee

\noindent The equation
for the $F_{(i)}$ is just the Schr\"{o}dinger equation
for a simple harmonic oscillator, whereas the
equation for the $G_{(i)}$ has an \emph{inverted} oscillator potential.  Of
course, this is the origin of the oscillatory and exponential behaviour
in the geodesics and zero modes of the string as discussed above.

Asymptotically, as $|x|,|y| \ra \infty$, we have
\be
F_{(i)}(x^i) \ra e^{\pm \frac{1}{2} \mu_{(i)} p_- (x^i)^2 }, \qquad
G_{(i)}(y^i) \ra
\frac{1}{\sqrt{y^i}} \cos \left( \frac{1}{2} \mu_{(i)} p_- (y^i)^2 \right),
\frac{1}{\sqrt{y^i}} \sin \left( \frac{1}{2} \mu_{(i)} p_- (y^i)^2 \right),
\ee

\noindent for each $F_{(i)}(x^i)$ and $G_{(i)}(y^i)$.  In the $x^i$
directions (with positive eigenvalues), the modes are either
exponentially damped, or are exponentially growing --- these are the
normalizable and non--normalizable modes of~\cite{leigh:02}.  In the
$y^i$ directions (with negative eigenvalues), however, the solutions
are oscillatory, the two signs giving left-- and right--moving plane
waves.  The former are easy to deal with: taking
\be
F_{(i)}(x^i) = e^{\pm \frac{1}{2} \mu_{(i)} p_-
(x^i)^2} f_{(i)}(x^i),
\ee

\noindent gives
\be
f_{(i)}'' \pm 2 \mu_{(i)} p_- x^i f_{(i)}' + (E_{(i)}\pm\mu_{(i)} p_-)
f_{(i)} = 0, 
\ee

\noindent which is of course solved in terms of Hermite polynomials,
up to an overall constant, giving a normalizable
solution~\cite{leigh:02,das:02} 
\be
F_{(i)}(x^i) = c_n e^{-\frac{1}{2} \mu_{(i)} p_- (x^i)^2}
H_n(\sqrt{\mu_{(i)} p_-} x^i), \qquad n = 
\frac{1}{2}\left( \frac{E_{(i)}}{\mu_{(i)} p_-} - 1 \right).
\ee

\noindent For real $n$, \emph{i.e.}, for real $E_{(i)}$, these solutions
furnish a complete basis of normalizable modes solving the equation of
motion in the $x^i$ directions.

We now turn to consider the
directions associated with negative eigenvalues.  Upon defining
$z^i=y^i/\alpha_{(i)}$, where $\alpha_{(i)}^2=1/(2\mu_{(i)}|p_-|)$,
the equation for the $G_{(i)}$ in (\ref{eqn:sep}) becomes 
\be
G_{(i)}''+ \left( \frac{1}{4} (z^i)^2 - a_{(i)} \right) G_{(i)} = 0,
\label{eqn:G}
\ee

\noindent where $a_{(i)} = -E_{(i)}\alpha_{(i)}^2$.  This can be
solved in terms of parabolic cylinder functions giving, for each
$z^i$, two independent solutions~\cite{abramowitz:64} 
\be
\psi_a(z) = W(a,z), \qquad \chi_a(z) = W(a,-z).
\label{eqn:cylinder}
\ee

\noindent The first solution has the limiting behaviour
\ba
\psi_a ( z \ra \infty ) &\ra& \sqrt{\frac{2k}{z}} \cos(z^2/4-a\log z
+\pi/4+\varphi/2), \\ 
\psi_a ( z \ra -\infty ) &\ra& \sqrt{\frac{2}{k|z|}}
\sin(z^2/4-a\log |z| +\pi/4+\varphi/2),
\ea

\noindent where $\varphi = \mbox{arg} \Gamma(\half+ia)$ and
$k=\sqrt{1+\exp(2\pi a)}-\exp(\pi a)$, and the second behaves as
\ba
\chi_a ( z \ra \infty ) &\ra& \sqrt{\frac{2}{kz}} \sin(z^2/4-a\log z
+\pi/4+\varphi/2), \\ 
\chi_a ( z \ra -\infty ) &\ra& \sqrt{\frac{2k}{|z|}} \sin(z^2/4-a\log
|z| +\pi/4+\varphi/2), 
\ea

To show that these functions are orthogonal, we write (\ref{eqn:G}) as
\be
G_a''+\frac{1}{4}z^2 G_a=a G_a,
\ee
     
\noindent where $G_a$ denotes either of the two functions
in (\ref{eqn:cylinder}), multiply both sides by $G_b$ and integrate by
parts, giving
\be
\int^L_{-L}G_a(z)G_b(z) ~\d z =
\frac{1}{(a-b)}\left[G_bG_a'-G_aG_b'\right]^L_{-L}. 
\ee

\noindent Taking the large $L$ limit and using the limiting behaviour
discussed above, we find
\be
\int^\infty_{-\infty} \psi_a(z) \psi_b(z) ~\d z =
\int^\infty_{-\infty} \chi_a(z) \chi_b(z) ~\d z = 2\pi\sqrt{1+e^{2\pi
    a}}~\delta(a-b).
\ee

\noindent Substituting instead for $G_a=\psi_a$ and $G_b=\chi_b$ gives
\be
\int^L_{-L}\psi_a(z) \chi_b(z) ~\d z = \frac{1}{(a-b)}
\left[\sqrt{\frac{k_b}{k_a}}-\sqrt{\frac{k_a}{k_b}}\right] \cos[(a-b)\log L].
\ee

\noindent For $a=b+\epsilon$, with $\epsilon$ small,
\be
k_a = k_b+\pi\epsilon\left[\frac{e^{2\pi b}}{\sqrt{1+e^{2\pi
	b}}}-e^{\pi b}\right], 
\ee

\noindent which shows that
\be
\int^\infty_{-\infty}\psi_a(z)\chi_b(z) ~\d z = \frac{\pi e^{\pi
    a}}{\sqrt{1+e^{2\pi a}}}\delta_{ab}, 
\ee

\noindent where $\delta_{ab}$ is the Kronecker delta.  We can then
write an arbitrary function as
\be
G(z)=\int \d a [\alpha(a)\psi_a(z)+\beta(a)\chi_a(z)]
\ee

\noindent where $\alpha$ and $\beta$ are found using the orthogonality
relations 
\be
\alpha(a)=\frac{1}{2\pi\sqrt{1+\exp(2\pi a)}}\int \d z \psi_a(z) G(z), \qquad
\beta(a)=\frac{1}{2\pi\sqrt{1+\exp(2\pi a)}}\int \d z \chi_a(z) G(z).
\ee

To recap, in the directions associated with negative eigenvalues we
found solutions in terms of cylinder functions and in the directions
associated with positive eigenvalues we found solutions in terms of
Hermite polynomials. If the parameters $a$ and $n$ are real then these
form a complete basis and we may use the orthogonality of these
functions to decompose any initial data set in terms of modes with
real $a$, $n$.  However, real $a$, $n$ implies real $E_{(i)}$ and with
it, from (\ref{eqn:Esum}), real $p_+$. That is, an arbitrary initial
perturbation profile is composed of modes with real $p_+$, and so the
system would seem to be stable.
     
To clarify the above, let us consider what happens in the more familiar
case of a free scalar field with negative (mass)$^2$:
\be
-\del_t^2\varphi+\del_x^2\varphi+\del_y^2\varphi+m^2\varphi=0.
\ee
     
\noindent To follow an analogue of the above argument we consider a
mode of momentum $p_x$,
\be
\varphi(t,x,y)=A(y)\cos(\omega t-p_xx)+B(y)\cos(\omega t-p_xx),
\ee
     
\noindent giving
\be
\del_y^2 A(y)=-(\omega^2+m^2-p_x^2)A(y)=-p_y^2A(y),
\ee

\noindent with a similar equation for $B(y)$.  The solutions to this
are either exponential or trigonometric, depending on the reality of
$p_y$ -- for normalizable solutions we need to take real momentum,
$p_y$, giving us trigonometric solutions. We are now able, using
standard Fourier analysis, to decompose an arbitrary initial
perturbation in terms of this basis of real $p_x$, $p_y$. This is the
analogue of being able to write solutions to (\ref{eqn:equation}) in
terms of cylinder functions with real $a$ and Hermite polynomials with
real $n$. The difference now is that $\omega^2=\underline{p}^2-m^2$,
so small momentum (long wavelength) states have imaginary $\omega$
whereas for the plane--wave we found that $p_+$ was always real.

One should be careful, however, since
there are subtleties associated with the lack of a true Cauchy
surface (some of which are discussed in~\cite{marolf:0210}), although the same conclusions
are reached in that work.  As we have mentioned above, it would seem
that evolving initial data from a $u=\mbox{constant}$ surface only misses
modes which are unlikely to exhibit unstable behaviour, since they are
unaffected by the wave.

%%%%%%%%%%%%%%%%%%%%%%%%%%%%%%%%%%%%%%%%%%%%%%%%%%%%%%%%%%%%%%%%%%%%%%%%

\sect{Conclusions}

The propagation of a closed string in the light--cone gauge on the
vacuum plane--wave gives rise to world--sheet scalars with a tachyonic
mass. In particular, the zero mode of the string, which determines the
supergravity modes, suffers from such a mass term; this motivated the
study of the classical stability of the supergravity solution.
Although such spacetimes are supersymmetric, and supersymmetry
is usually understood to imply stability, the standard arguments
do not apply as the metrics are not asymptotically flat (they
are Lorentzian symmetric spaces). It may turn out however that
such methods could be modified to cover the plane--wave, and this would
be an interesting avenue of research.
     
Taking the IIB equations of motion, perturbed around the vacuum
plane wave, we have shown that in a gauge suited to the spacetime
(light--cone gauge) the system boils down to the study of
a single equation, (\ref{eqn:equation}). The simplicity of this
equation allowed us to find its general solution in terms of 
known functions. Moreover, an arbitrary initial perturbation
was seen to be decomposable in terms of modes with real $p_+$,
corresponding to oscillatory motion, and thereby providing strong
evidence for the stability of the spacetime.

Of course, one would like to be able to \emph{prove} that instabilities
do not exist, by showing that there are no normalizable solutions of
(\ref{eqn:equation}) with imaginary $p_+$.  Unfortunately, however, we
have not been able to do this.

On general grounds, one might suspect that the presence of tachyonic
worldsheet scalars would not give rise to any instabilities of the
background, for the simple reason that the mass of these scalars is
purely an artifact of choosing the light--cone gauge on the
worldsheet~\cite{gimon:02,cvetic:0209}.  Choosing instead, for example, the
static gauge, gives rise to a worldsheet theory of massless but
interacting scalars~\cite{cvetic:0209}.  Indeed, simply changing
from Brinkman to Rosen coordinates in the background spacetime has a
similar effect on the worldsheet.  It would be interesting to
investigate how the interactions in different coordinates or gauges
mimic the effect of a mass, tachyonic or otherwise.

%%%%%%%%%%%%%%%%%%%%%%%%%%%%%%%%%%%%%%%%%%%%%%%%%%%%%%%%%%%%%%%%%%%%%%%%%

\vspace{1cm}
\noindent
{\bf Acknowledgements}\\
We are very grateful to Don Marolf for many enlightening discussions,
and for providing us with an early copy of the
preprint~\cite{marolf:0210}, and to Simon Ross for a critical reading
of this manuscript.  DB would
further like to thank Rob Myers and Moshe Rozali for interesting
conversations, and is supported in part by NSERC.  PMS is supported by a PPARC fellowship.

\end{document}